# On inertia forces and cosmological acceleration

Yurii A. Spirichev

The State Atomic Energy Corporation ROSATOM, "Research and Design Institute of Radio-Electronic Engineering" - branch of Federal Scientific-Production Center "Production Association "Start" named after Michael V.Protsenko", Zarechny, Penza region, Russia

E-mail: yurii.spirichev@mail.ru
05.02.2022 г.

**Abstrakt:** It is shown that the equations of gravitational waves and cosmological acceleration follow from the tensor of inertial forces. The possibility of the existence of three types of gravitational waves: transverse, potential and vortex waves are shown. The equation of "free cosmological fall" of matter is obtained. It is concluded that the cosmological acceleration is an attribute of the geometry of space-time and for its explanation does not require the involvement of the hypothesis of "dark" energy.

**Contents**
1 Introduction
2 The tensor of inertial forces of a material point, equations of motion and cosmological acceleration
3 Gravitational waves and the generalized continuity equation in pseudo-Euclidean space-time
4 Conclusion
References

### 1. Introduction

An important problem of modern cosmology is the explanation of the cause of the accelerated expansion of the universe discovered in 1998. [1 - 3]. Within the standard cosmological model of Lambda Cold Dark Matter (λCDM), based on the General Theory of Relativity (GTR), this phenomenon is explained by the hypothesis of the existence of "dark" energy, which creates a constant negative pressure in space and causes the accelerated expansion of the universe. So far the nature or carrier of this "dark" energy has not been determined, although according to the existing calculations it makes up about 69% of all energy of the universe [4]. To explain such a huge contribution of "dark" energy to the total energy balance, a number of hypotheses have been developed about the existence in nature of an additional scalar field, responsible for this energy [5 - 8], etc. This scalar field should manifest itself not only in the form of cosmological expansion, but no such additional interaction with matter has been detected.

D. Ivanenko noted in the introductory article to the book by J. Weber [9] in cosmology there is "empirically fair", but "rather surprising" relation $GM^2/R \approx Mc^2$ [10] of approximate equality of the gravitational energy of the total mass of the universe M and rest energy of ordinary matter for its apparent size R at that time. These words were written before the discovery of cosmological acceleration and the creation of the "dark" energy hypothesis. Obviously, the "dark" energy hypothesis does not fit into this "empirically fair" relation of equality of rest energy and gravitational energy of visible matter in the universe.

Another important problem of cosmology is gravitational waves, the existence of which follows from the equations of GTR. After a long search in 2015 gravitational waves were reliably detected and now their research is under way. Since the equivalence of the forces of inertia, gravitation and curvature of space-time geometry is accepted in GTR, the gravitational waves are understood as propagating perturbations of the pseudo-Riemannian space-time metric, having a force character. On this occasion E. Schrödinger wrote [11] that in GTR "the fundamental principle of equivalence of acceleration and gravitational field, clearly means that there is no place here for any "forces" producing acceleration, except gravitation, which, however, should not be considered as a force, but as a property of space-time geometry". In GTR the equation of gravitational waves is obtained from Einstein's equation, the right side of which is an energy-momentum tensor including energy of matter and energy of gravitational field. In the simplest method of obtaining the wave equation, the condition of empty space and the smallness of the gravitational field value is taken. The metric tensor is taken as pseudo-Euclidean and the tensor of small perturbations, which create a small gravitational field, is added to it. Then, by transforming the reference frame, using calibration and imposing additional conditions, a wave equation for the tensor of small perturbations, i.e. the gravitational field in the void, is obtained [11].

Besides this there is a number of more complicated methods for obtaining the wave equation of the gravitational field, some of which are described in [12] - [14]. However, as F. Pirani [14] notes, the methods using weak field approximations and the choice of mathematically convenient coordinate conditions and calibrations are disputable, and their physical meaning is not clear. Thus, the derivation of the equations of gravitational waves by other, mathematically more transparent methods is still actual.

The equations of motion of matter and gravitational waves in the pseudo-Riemannian space follow from the Einstein equation describing the energy relation of matter and space-time geometry. However, it is possible to approach consideration of these questions not only on the basis of the energy approach, but also on the basis of the force approach, i.e. to obtain the equations of motion of matter we use not the energy-momentum tensor and Lagrange formalism, but the tensor of inertial forces acting on matter. This tensor of inertial forces is the four-dimensional derivative of the four-dimensional momentum of matter.

Another problem of cosmology is the explanation of the nature of cosmological jets observed at the nuclei of some active galaxies, protoplanets, quasars, and black holes [15-17]. The main hypothesis of the formation of jets is considered to be the motion of electric charges of plasma in the magnetic field. However, plasma is quasi-neutral and there are no convincing reasons for violation of this quasi-neutrality in large regions of space to create huge currents and the corresponding magnetic field

Since in our universe the nonlinear pseudo-Riemannian space differs from the linear pseudo-Euclidean space by not more than 1% and locally is reduced to the Minkowski space, in the present paper the matter motions will be considered in the pseudo-Euclidean Minkowski space. The velocities of cosmological motions of matter, in our part of the universe, are far from relativistic, so the equations of motion here will be considered in the nonrelativistic approximation. In GTR the principle of equivalence of gravitation and inertia is valid, and the inertial motion of matter in the gravitational field can be curvilinear and accelerated. It follows from this that any free motion of matter in pseudo-Euclidean space, on which only the forces of inertia and gravitation act, is an inertial motion.

The purpose of this paper is to obtain and consider the equations of inertial motion of matter in the pseudo-Euclidean Minkowski space in the nonrelativistic approximation and to derive the equations of gravitational waves from the four-dimensional tensor of inertial forces. Since the inertial forces are equivalent to the gravitational forces, this approach to obtaining the equations of gravitational waves allows to do without transformations of the reference systems, calibrations, additional assumptions and restrictions. In addition, it allows us to obtain new information concerning the physics of gravitational waves.

In this paper, four-dimensional vector quantities have a representation in a fixed reference frame and an orthogonal linear coordinate system with real temporal and imaginary spatial components [18]. Four-dimensional coordinates of a material point are defined as a four-dimensional radius-vector $\mathbf{R}_\nu(c \cdot t, i \cdot \mathbf{r})$, where t is time, $\mathbf{r}$ is a three-dimensional radius-vector. In the special theory of relativity (STR) for a free-moving material point the main relations are [19]:

$$E^2 - p^2 = m^2 \cdot c^4 \qquad \text{and} \qquad \mathbf{p} = E \cdot \mathbf{v}/c^2$$

Here E is the energy of a material point, $\mathbf{p}$ is the vector of its momentum, $\mathbf{v}$ is the vector of its velocity. From these relations follow the expressions for energy and momentum in the form [19]:

$$E = m \cdot c^2 / \sqrt{1 - v^2/c^2} \qquad \text{and} \qquad \mathbf{p} = m \cdot \mathbf{v} / \sqrt{1 - v^2/c^2}$$

Energy E and momentum $\mathbf{p}$ are components of the four-dimensional momentum Energy E and momentum p are components of the four-dimensional momentum $\mathbf{P}_\nu(E/c, i \cdot \mathbf{p})$ or $\mathbf{P}_\nu(m \cdot c \cdot \gamma, i \cdot m \cdot \mathbf{v} \cdot \gamma)$, where $\gamma = 1/\sqrt{1 - v^2/c^2}$ is the relativistic coefficient. According to modern concepts, mass is an invariant quantity, independent of the velocity of the body, then the four-dimensional momentum can be written in the form $\mathbf{P}_\nu(m \cdot \mathbf{V}_\nu(\gamma \cdot c, i \cdot \gamma \cdot \mathbf{v}))$, where $\mathbf{V}_\nu(\gamma \cdot c, i \cdot \gamma \cdot \mathbf{v})$ is the four-dimensional velocity vector. Для принятого представления псевдоевклидовой геометрии пространства-времени, оператор четырехмерной частной производной имеет вид For the accepted representation of the pseudo-Euclidean space-time geometry, the four-dimensional partial derivative operator has the form $\partial_\mu(\partial_t/c, -i\cdot\nabla)$. For the accepted representation of this operator and four-dimensional vector quantities, it is possible not to distinguish between covariant and contravariant indices of four-dimensional vectors and tensors.

## 2. The tensor of inertial forces of a material point, equations of motion and cosmological acceleration

Methods for obtaining wave equations based on energy-momentum relations contain certain mathematical difficulties. This is due to the fact that from the energy-momentum tensor, by definition, follow conservation equations that describe a closed system in which the energy-momentum is conserved. They do not describe wave motions, when the system is open and the energy-momentum is transferred out of it by wave motion. Let us find the equations of inertial motion for an open system from the tensor of inertial forces. For this purpose, we obtain the relativistic tensor of inertial forces in the form of a four-dimensional derivative of the relativistic momentum:

$$F_{\mu\nu}^R = \partial_\mu \mathbf{P}_\nu^R(\gamma\cdot m\cdot c, i\cdot\gamma\cdot\mathbf{p}) = \mathbf{P}_\nu \partial_\mu \gamma + \gamma\cdot\partial_\mu \mathbf{P}_\nu \qquad (1)$$

Here $\partial_\mu \gamma = (\partial_t \gamma/c, -i\cdot\nabla\gamma)$ is a four-dimensional vector. For the nonrelativistic approximation, when $\gamma \approx 1$, the tensor of inertial forces has the form:

$$F_{\mu\nu} = \partial_\mu P_\nu = \begin{pmatrix} \partial_t m & \frac{1}{c}i\cdot\partial_t p_x & \frac{1}{c}i\cdot\partial_t p_y & \frac{1}{c}i\cdot\partial_t p_z \\ -i\cdot c\cdot\partial_x m & \partial_x p_x & \partial_x p_y & \partial_x p_z \\ -i\cdot c\cdot\partial_y m & \partial_y p_x & \partial_y p_y & \partial_y p_z \\ -i\cdot c\cdot\partial_z m & \partial_z p_x & \partial_z p_y & \partial_z p_z \end{pmatrix} \qquad (2)$$

Taking into account the constancy of the mass of a material point, we obtain the tensor of inertial forces in the form:

$$F_{\mu\nu} = \partial_\mu P_\nu = \begin{pmatrix} 0 & \frac{1}{c}i\cdot\partial_t p_x & \frac{1}{c}i\cdot\partial_t p_y & \frac{1}{c}i\cdot\partial_t p_z \\ 0 & \partial_x p_x & \partial_x p_y & \partial_x p_z \\ 0 & \partial_y p_x & \partial_y p_y & \partial_y p_z \\ 0 & \partial_z p_x & \partial_z p_y & \partial_z p_z \end{pmatrix} \qquad (3)$$

In the absence of external forces, we obtain the equations of inertial motion of a material point in the form of divergence of the tensor (3), which we equate to zero. Since the tensor (3) is asymmetric and its divergences by each index are different, it is necessary to take the sum of divergences by each index:

$$\partial_\mu F_{\mu\nu} + \partial_\nu F_{\mu\nu} = 0 \qquad (4)$$

Let us write this equation of inertial motion of a material point as a system of three-dimensional equations:

$$\nabla \cdot \partial_t \mathbf{p}/c = 0 \qquad (5)$$

$$(\frac{1}{c^2}\partial_{tt}\mathbf{p} - \Delta \mathbf{p}) - \nabla(\nabla \cdot \mathbf{p}) = 0 \qquad (6)$$

Taking the time derivative of Eq. (6) and taking into account Eq. (5), we obtain the wave equation for the Newtonian force of inertia:

$$(\frac{1}{c^2}\partial_{tt} - \Delta)\partial_t \mathbf{p} = 0 \qquad (7)$$

From Eq. (7) mass can be excluded and write the wave equation in the form:

$$(\frac{1}{c^2}\partial_{tt} - \Delta)\partial_t \mathbf{v} = (\frac{1}{c^2}\partial_{tt} - \Delta)\mathbf{g} = 0 \qquad (8)$$

Here **g** is the acceleration of a material point. Since the forces of inertia are equivalent to gravity, the wave Eq. (7) for the Newtonian force of inertia and acceleration (8) can be viewed as equations of gravitational waves propagating with the speed of light. It follows from Eq. (5) that in the absence of external forces, the divergence of the Newtonian force of inertia for a material point is zero. By integrating Eq. (5) over space, we obtain:

$$\partial_t \mathbf{p} = \mathbf{F} \text{ - const at } \mathbf{r} \qquad (9)$$

It follows from this equation that at any point of pseudo-Euclidean space-time there is a constant Newtonian force of inertia **F** acting on a material point of mass m. The constant mass can be excluded from Eq. (5) and after integrating it, the equation can be written in the form:

$$\partial_t \mathbf{v} = \mathbf{a} \text{ - const at } \mathbf{r} \qquad (10)$$

It follows that at any point of pseudo-Euclidean space-time for any mass there is a constant acceleration **a**. This conclusion corresponds to the existence of cosmological acceleration independent of the mass of galaxies. By integrating Eq. (5) over time, we obtain:

$$\nabla \cdot \mathbf{p} = D \text{ - const at t} \quad \text{or} \quad m \cdot \nabla \cdot \mathbf{v} = D \text{ - const at t} \qquad (11)$$

It follows from this equation that at any point in space there is a divergence of momentum constant in time. This is Gauss's law for momentum, where D is a constant in time scalar source of momentum. Since a constant mass can be excluded from Eq. (5), there is a constant velocity divergence at every point in space.

From Eq-s (5) - (11) the conclusion follows about the existence of gravitational waves and cosmological accelerated expansion of the universe. This effect does not depend on the mass of matter and is a property of the space-time geometry. The conclusions given here are made in nonrelativistic approximation for linear pseudo-Euclidean space-time, but they are fundamentally valid for the relativistic case and nonlinear pseudo-Riemannian space. It follows that cosmological accelerated expansion of the Universe is an attribute of pseudo-Euclidean and pseudo-Riemannian geometry of spacetime and for explanation of this phenomenon the hypothesis of existence of "dark" energy is not required. Eq. (9) says that a material point is subject to a constant Newtonian force. But Newton's force of inertia is a counter force to another gravitational force. This gravitational force will be discussed in the next section.

## 3. Gravitational waves and the generalized continuity equation in pseudo-Euclidean space-time

Let us consider a more general case in the form of motion of homogeneous matter with mass density σ in pseudo-Euclidean space-time. For this purpose, in tensors (1) and (2) and in Eq. (4) we will pass from a constant mass m of a material point to a variable mass density σ and, respectively, to the momentum density $\mathbf{p} = \sigma \cdot \mathbf{v}$. For this case, let us write the equation of inertial motion of matter (4) in the form of a system of three-dimensional equations:

$$c \cdot (\partial_{tt}\sigma/c^2 - \Delta\sigma) = -\partial_t(\partial_t\sigma + \nabla \cdot \mathbf{p})/c \qquad (12)$$

$$\partial_{tt}\mathbf{p}/c^2 - \Delta\mathbf{p} = \nabla(\partial_t\sigma + \nabla \cdot \mathbf{p}) \qquad (13)$$

The left parts of these equations have a wave character, and the right parts represent the derivatives of the invariant $\partial_t\sigma + \nabla \cdot \mathbf{p}$ of the inertial force tensor. This invariant is the 4-divergence of the 4-momentum $\mathbf{P}_\nu$. If this invariant is equal to zero, it passes into the canonical Euler continuity equation $\partial_\nu \mathbf{P}_\nu = \partial_t\sigma + \nabla \cdot \mathbf{p} = 0$. In the case under consideration the invariant is not equal to zero, hence, the derivatives of the 4-divergence of the 4-pulse $\mathbf{P}_\nu$ can be considered as a source of waves described by the left-hand sides of Eq.-s (12) and (13), and the equations themselves can be written in the form:

$$(\partial_{tt}/c^2 - \Delta)(-\sigma \cdot c^2) = \partial_t(\partial_\nu \mathbf{P}_\nu) \qquad (14)$$

$$\partial_{tt}\mathbf{p}/c^2 - \Delta\mathbf{p} = \nabla(\partial_\nu \mathbf{P}_\nu) \qquad (15)$$

These equations can be seen as a generalization of the Euler continuity equation to the wave case. Let us take the gradient from both parts of Eq. (14) and the time derivative from both parts of Eq. (15):

$$(\partial_{tt}/c^2 - \Delta)(-\nabla\sigma \cdot c^2) = \nabla\partial_t(\partial_\nu \mathbf{P}_\nu) \qquad (16)$$

$$(\partial_{tt}/c^2 - \Delta)(\partial_t\mathbf{p}) = \partial_t\nabla(\partial_\nu \mathbf{P}_\nu) \qquad (17)$$

The left-hand wave parts of Eq.-s (16) and (17) describe waves of two kinds of inertial forces:

$$(\partial_{tt}/c^2 - \Delta)\nabla(\sigma \cdot c^2) = 0 \qquad (18)$$

$$(\partial_{tt}/c^2 - \Delta)\partial_t\mathbf{p} = 0 \qquad (19)$$

Eq. (18) includes the potential force of the rest mass energy density gradient of the unit volume of the medium $\nabla(\sigma \cdot c^2)$ or $\nabla(m \cdot c^2)$. Eq. (19) includes the Newtonian inertial force density. Since the forces of inertia are equivalent to gravity, the energy expression $\sigma \cdot c^2$ can be viewed as the energy density of gravity. Then the square of the speed of light can be regarded as the gravitational potential, i.e. $c^2 = \varphi$. This conclusion also follows from the known empirical relation $GM^2/R \approx Mc^2$ [10] for the energy density of the total mass of the universe M. This relation can be transformed by reducing the mass of M in its both parts and obtain $GM/R \approx c^2$ or $\varphi \approx c^2$. Hence, it follows that the known formula for the rest energy of matter $E = m \cdot c^2$ describes potential energy of matter in the field of this gravitational potential.

Given the equality of the right-hand sides of Eq.-s (18) and (19), we can write down the equality of their left-hand wave parts:

$$(\partial_{tt}/c^2 - \Delta)(-\nabla \sigma \cdot c^2) = (\partial_{tt}/c^2 - \Delta)(\partial_t \mathbf{p}) \text{ or } (\partial_{tt}/c^2 - \Delta)(-\nabla \sigma \cdot c^2 - \partial_t \mathbf{p}) = 0 \qquad (20)$$

The trivial solution of this equation is equality $-\nabla(\sigma \cdot c^2) = \partial_t \mathbf{p}$. This equality is analogous to the well-known equation $-\nabla(m \cdot \varphi) = \partial_t \mathbf{p}$ of free fall of a body of mass m in the field of gravitational potential φ. Consequently, the accelerated cosmological expansion of the universe can be considered as a free fall of matter in the field of gravitational potential $\varphi = c^2$. Then the left part of this equation describes the potential force of gravitation and the right part the Newtonian force of inertia. Taking the divergence of both parts of this equality, we obtain $-\Delta(\sigma \cdot c^2) = \nabla \cdot \partial_t \mathbf{p}$. It follows that the source of the Newtonian force of inertia is the value of $-\Delta(\sigma \cdot c^2)$. Thus, the force leading to the expansion of the universe is the gravitational potential force $\mathbf{F}_G = -\nabla(\sigma \cdot c^2)$, where σ is the average density of matter of the universe.

Taking the gradient from both parts of the relation $GM^2/R \approx Mc^2$ and replacing in it the matter mass M by its average density σ, we obtain $\nabla(G\sigma^2/R) \approx \nabla(\sigma \cdot c^2)$. Then, replacing its right part, we can write $\mathbf{F}_G = \sigma \cdot \mathbf{a} \approx -\nabla(G\sigma^2/R)$, where $\mathbf{a}$ is the acceleration of the " free cosmological fall". Then the cosmological acceleration is $\mathbf{a} \approx -\dfrac{1}{\sigma}\nabla(G\sigma^2/R)$.

At "free cosmological fall" $-\nabla(\sigma \cdot c^2) = \partial_t \mathbf{p}$ there is no gravitational radiation, then the wave parts of Eq.-s (16) and (17) are equal to zero:

$$\partial_t \nabla(\partial_v \mathbf{P}_v) = 0 \qquad (21)$$

By integrating this equation over three-dimensional space, we obtain the equation:

$$\partial_t(\partial_v \mathbf{P}_v) = U \text{ - const at } \mathbf{r} \qquad (22)$$

It follows that the rate of change in time of the 4-divergence of the density 4-momentum $\mathbf{P}_v$ at any point in space is a constant value. By integrating Eq. (21) over time, we obtain:

$$\nabla(\partial_v \mathbf{P}_v) = \mathbf{H} \text{  - const at t} \qquad (23)$$

It follows from this equation that the three-dimensional gradient 4-divergence of the density 4-momentum $\mathbf{P}_\nu$ is a constant in time. It follows from Eq.-s (21) - (23) that, in the absence of external forces, there is a constant 4-momentum density $\mathbf{P}_\nu$ flux from any closed three-dimensional volume of space with matter. This suggests that in pseudo-Euclidean (pseudo-Riemannian) space-time the energy-momentum density in any closed three-dimensional volume will continuously decrease, which corresponds to the observed cosmological expansion of the universe.

Let us take the rotor from both parts of Eq. (17) and obtain the wave equation:

$$(\partial_{tt}/c^2 - \Delta)(\nabla \times \partial_t \mathbf{p}) = 0 \qquad (24)$$

This equation describes the vortex waves of Newtonian inertial force. It follows from this equation that the accelerated rotational motion of matter leads to the emission of vortex waves of Newtonian force of inertia oriented along the rotation axis. These waves of inertial force, propagating along the rotation axis of the medium, interact with the matter and entrain the rotating matter. This may explain cosmological jets observed from the nuclei of some active galaxies, protoplanets, quasars, and black holes. The presence of jets indicates the accelerated rotation of their central regions as a result of gravitational compression.

### 4. Conclusion

At present, the equations of motion of matter in GTR are derived from geometric equations using the Lagrange formalism and the energy-momentum tensor, i.e. based on the balance of energy relations for a closed system in which conservation of energy-momentum is assumed. In the present article, to obtain the equations of motion of matter, the inertial force tensor (2) is used in the form of a four-dimensional derivative of the 4-momentum, while the conservation of energy-momentum is not assumed in the system. The divergence of this tensor is the equations of motion of matter for an open system, from which energy-momentum can be radiated in the form of waves. These equations can be considered a generalization of the Euler continuity equation to the wave case. From the equations of motion of matter it follows that the derivatives of the 4-divergence 4-momentum are the source of three kinds of gravitational waves: transverse waves of the Newtonian force of inertia (19), potential waves of the rest energy gradient of matter (18) and vortex waves of the Newtonian force of inertia (24). It follows from equations of motion (9) - (11) and (21) - (23) that at every point of pseudo-Euclidean (pseudo-Riemannian) space-time a constant Newtonian force of inertia acts on matter, and matter has a constant acceleration, observed as an accelerated cosmological expansion of the universe. This cosmological expansion obeys the equation of free fall of matter in the field of gravitational potential $-\nabla(\sigma \cdot c^2) = \partial_t(\sigma \cdot \mathbf{v})$, which is numerically equal to the square of the speed of light. Thus, the force leading to the expansion of the universe is the gravitational potential force $\mathbf{F}_G = -\nabla(\sigma \cdot c^2)$, where σ is the average density of matter of the universe, and $c^2$ is the gravitational potential.

## 5. References

1. Riess A. G. et al. Observational evidence from supernovae for an accelerating universe and a cosmological constant. Astron. J. 116, 1009–1038. 1998.
2. Schmidt B. P. et al. The High Z supernova search: Measuring cosmic deceleration and global curvature of the universe using type Ia supernovae. Astrophys. J. 507, 46–63. 1998.
3. Perlmutter S. et al. Measurements of Omega and Lambda from 42 high redshift supernovae. Astrophys. J. 517, 565–586. 1999.
4. Ade P. A. R. et al. Planck 2015 results. XIII. Cosmological parameters. Astron. Astrophys. 594, A13. 2016.
5. Zlatev I., Wang L.-M. and Steinhardt P. J. Quintessence, cosmic coincidence, and the cosmological constant. Phys. Rev. Lett. 82, 896–899. 1999.
6. Caldwell R. R. and Kamionkowski M. The Physics of Cosmic Acceleration. Ann. Rev. Nucl. Part. Sci. 59, 397–429. 2009.
7. Chang H.-Y. and Scherrer R. J. Reviving Quintessence with an Exponential Potential. arXiv:1608.03291. 2016.
8. Dutta S. and Scherrer R. J. Dark Energy from a Phantom Field Near a Local Potential Minimum. Phys. Lett. B676, 12–15. 2009.
9. Weber J. General Theory of Relativity and Gravitational Waves Moscow: Foreign Literature, 1962.
10. Feynman R.P., Morinigo F.B., Wagner W.G., Hatfield B. Feynman Lectures on Gravitation. Addison-Wesley, Reading, MA, 1995.
11. Schrödinger E. Spacetime structure of the Universe Translated into Russian, Moscow: Nauka, 1986.
12. Landau L D, Lifshitz E M The Classical Theory of Fields (Oxford: Butterworth-Heinemann, 2000)
13. Fok V.A. Theory of Space, Time and Gravitation. MOSCOW: GITL 1956
14. Pirani F. The invariant formulation of the theory of gravitational radiation Phys. Rev., 105, 1089—1099 1957
15. Roger D. Blandford and Roman L. Znajek Electromagnetic extractions of energy from Kerr black holes. Mon. Not. Roy. Astron. Soc. 179, 433 - 456 1977.
16. Masaaki Kusunose and Fumio Takahara A photo-hadronic model of the Large scale jets of 3C 273 and PKS 1136–135 arXiv: 1805.10764 2018.
17. Jacobson T. and Rodriguez M. J. Blandford-Znajek process in vacuo and its holographic dual arXiv: 1709.10090 2017.
18. Tonnela M.-A. Fundamentals of electromagnetism and the theory of relativity, in Russian, Moscow: Foreign Literature, 1962
19. Okun L.B. Mass. Energy. Relativity. UFN vol.158 vol.3 1989, pp. 511-530